# Recursive Information Hiding in Visual Cryptography

Sandeep Katta

Department of Computer Science, Oklahoma State University

Stillwater, OK 74078

**Abstract:** Visual Cryptography is a secret sharing scheme that uses the human visual system to perform computations. This paper presents a recursive hiding scheme for *3* out of *5* secret sharing. The idea used is to hide smaller secrets in the shares of a larger secret without an expansion in the size of the latter.

## 1. INTRODUCTION

Secret sharing was invented independently by G.R. Blakley and Adi Shamir [2, 12]. It was generalized to visual cryptography by Moni Naor and Adi Shamir [1]. The basic model for visual sharing of the *k* out of *n* secret image is such that;

- Any *n* participants can compute the original message if any *k* (or more) of them are stacked together.
- No group of *k-1* (or fewer) participants cannot compute the original message.

Images are split into two or more shares such that when a predetermined number of shares are aligned and stacked together, then the secret image is revealed [1, 3] without any computation.

Recursive hiding of secrets is proposed in [4, 5, 16-18]. The idea involved is recursive hiding of smaller secrets in shares of larger secrets with secret sizes increasing at every step. While the scheme presented in [4] is a non-threshold scheme, schemes in [5, 16-19] are threshold schemes. In [16], a tree data structure is used for recursive encoding of text such that the internal nodes of the tree also carry information in addition to leaves of the tree. Repeated application of Shamir's scheme is used in [17] to share *k-1* secrets in *n* shares. However, the general threshold recursive schemes in [16-19] are not visual cryptography schemes. The scheme in [5], although a recursive visual cryptography scheme, is restricted to *2-out-of-n* scheme. In this article, we take a step further and present a *3-out-of-5* recursive scheme for visual secret sharing.

Further this idea can be generalized to a *3* out of '*n*' threshold scheme. We consider only monochrome images, and each pixel (or sub-pixel) is considered to be one bit of information.

## 2. BASIC MODEL

In visual secret sharing, the message bit consists of a collection of black and white pixels and each pixel is handled separately.

- Each pixel in the original image appears in "*n*" modified versions, one for each transparency and they are called *shares*.
- Each share is a collection of "*m*" black and white subpixels.

The resulting picture can be thought as a [$n \times m$] Boolean matrix S = [$s_{ij}$]

- $s_{ij}$ = 1 if the $j^{th}$ subpixel in the $i^{th}$ share is black.
- $s_{ij}$ = 0 if the $j^{th}$ subpixel in the $i^{th}$ share is white.

**Preliminary Notation**

- **$n$**     →     Group Size
- **$k$**     →     Threshold
- **$m$**     →     Pixel Expansion
- **$\alpha$**     →     Relative Contrast
- **$C_0$**     →     Collection of $n \times m$ Boolean matrices for shares of White pixel
- **$C_1$**     →     Collection of $n \times m$ Boolean matrices for shares of Black pixel
- **$V$**     →     OR'ed $k$ rows
- **$H(V)$**     →     Hamming weight of $V$
- **$d$**     →     number in [1,$m$]
- **$r$**     →     Size of collections $C_0$ and $C_1$

**Definition 1.** A solution to the *k* out of *n* visual secret sharing scheme consists of two collections of *n* x *m* Boolean matrices $C_0$ and $C_1$. To share a white pixel, one of the matrices in $C_0$ is randomly chosen and to share a black pixel, one of the matrices in $C_1$ is randomly chosen. The chosen matrix defines the color of the *m* subpixels in each one of the *n* transparencies. The solution is considered valid if the following three conditions are met [1].

**Contrast**
- For S in $C_0$ (WHITE): $H(V) \leq d - \alpha m$
- For S in $C_1$ (BLACK): $H(V) \geq d$

**Security**
- The two collections of $q \times m$ (*1≤q≤k*) matrices, formed by restricting $n \times m$ matrices in $C_0$ and $C_1$ to any $q$ rows, are indistinguishable.

**Properties of "*3*" out of "*n*" scheme, (*n* ≥ 3)**

- Pixel Expansion, $m = 2n - 2$.
- Relative Contrast, $\alpha = \frac{1}{2n} - 2$.
- Let B be the black $n \times (n-2)$ matrix which contains only 1's.
- Let I be the Identity $n \times n$ matrix which contains 1's on the diagonal and 0's elsewhere.
- BI is an $n \times (2n-2)$ concatenated matrix.
- $c$(BI) is the complement of BI.
- $C_0$ contains matrices obtained permuting columns of $c$(BI).
- $C_1$ contains matrices obtained permuting columns of BI.
- Any single share contains an arbitrary collection of *(n-1)* black & *(n-1)* white subpixels.
- Any pair of shares has *(n-2)* common black & two Individual black subpixels.
- Any stacked triplet of shares from $C_0$ has *n* black subpixels.
- Any stacked triplet of shares $C_1$ has *(n+1)* black subpixels.

Based upon the following properties we can design the matrix for "3" (*k*) out of "5" (*n*) scheme.

B = $n \times (n-2) \rightarrow 5 \times (5-2) = 5 \times 3$

$$\begin{bmatrix} 1 & 1 & 1 \\ 1 & 1 & 1 \\ 1 & 1 & 1 \\ 1 & 1 & 1 \\ 1 & 1 & 1 \end{bmatrix}$$
*5 × 3*

I = $n \times n = 5 \times 5$

$$\begin{bmatrix} 1 & 0 & 0 & 0 & 0 \\ 0 & 1 & 0 & 0 & 0 \\ 0 & 0 & 1 & 0 & 0 \\ 0 & 0 & 0 & 1 & 0 \\ 0 & 0 & 0 & 0 & 1 \end{bmatrix}$$
*5 × 5*

Two collections of $n \times m$ Boolean matrices $C_0$ and $C_1$.

- $m = 2n - 2 \rightarrow 2(5) - 2 = 8$
- *5 × 8* Boolean matrices.

$$c(\text{BI}) = \begin{bmatrix} 0 & 0 & 0 & 0 & 1 & 1 & 1 & 1 \\ 0 & 0 & 0 & 1 & 0 & 1 & 1 & 1 \\ 0 & 0 & 0 & 1 & 1 & 0 & 1 & 1 \\ 0 & 0 & 0 & 1 & 1 & 1 & 0 & 1 \\ 0 & 0 & 0 & 1 & 1 & 1 & 1 & 0 \end{bmatrix} \begin{matrix} \rightarrow & \text{Share1} \\ \rightarrow & \text{Share2} \\ \rightarrow & \text{Share3} \\ \rightarrow & \text{Share4} \\ \rightarrow & \text{Share5} \end{matrix}$$
*5 × 8*

[1]   [2]   [3]   [4]   [5]   [6]   [7]   [8]

Hamming weight of $c$(BI) i.e. is White $H(V) = 5$

$$BI = \begin{bmatrix} 1 & 1 & 1 & 1 & 0 & 0 & 0 & 0 \\ 1 & 1 & 1 & 0 & 1 & 0 & 0 & 0 \\ 1 & 1 & 1 & 0 & 0 & 1 & 0 & 0 \\ 1 & 1 & 1 & 0 & 0 & 0 & 1 & 0 \\ 1 & 1 & 1 & 0 & 0 & 0 & 0 & 1 \end{bmatrix} \begin{matrix} \rightarrow & \text{Share1} \\ \rightarrow & \text{Share2} \\ \rightarrow & \text{Share3} \\ \rightarrow & \text{Share4} \\ \rightarrow & \text{Share5} \end{matrix}$$

$5 \times 8$

[1]   [2]   [3]   [4]   [5]   [6]   [7]   [8]

Hamming weight of BI i.e. is Black $H(V) = 8$

$C_0$ = {all the matrices obtained by permuting the columns of c(BI)}

$C_1$ = {all the matrices obtained by permuting the columns of BI}

If the columns are not permuted then there is a possibility to reveal the secret information in any single share and therefore the process fails.

For example here are few different permutations.

{[1]   [8]   [2]   [7]   [3]   [6]   [4]   [5]}

{[2]   [4]   [6]   [8]   [1]   [3]   [5]   [7]}

{[3]   [2]   [1]   [8]   [7]   [6]   [4]   [5]}

{[5]   [8]   [1]   [6]   [2]   [3]   [7]   [4]}

Shares of a white Pixel

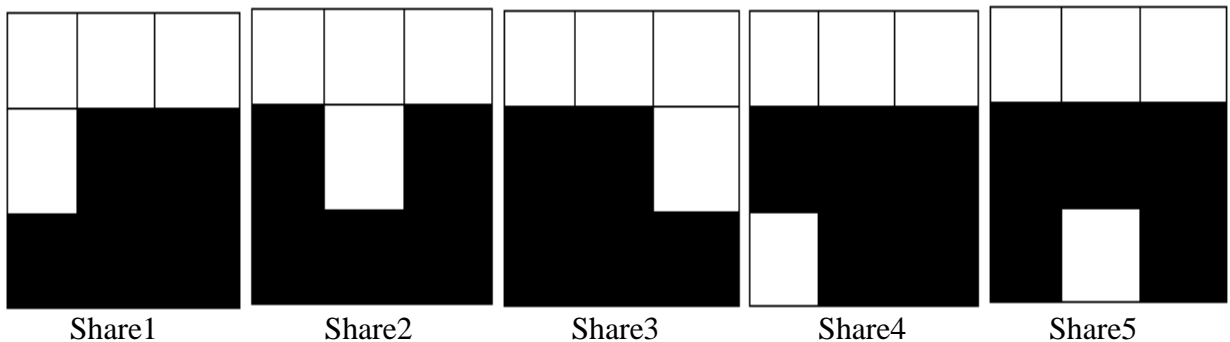

Share1          Share2          Share3          Share4          Share5

Shares of a Black Pixel

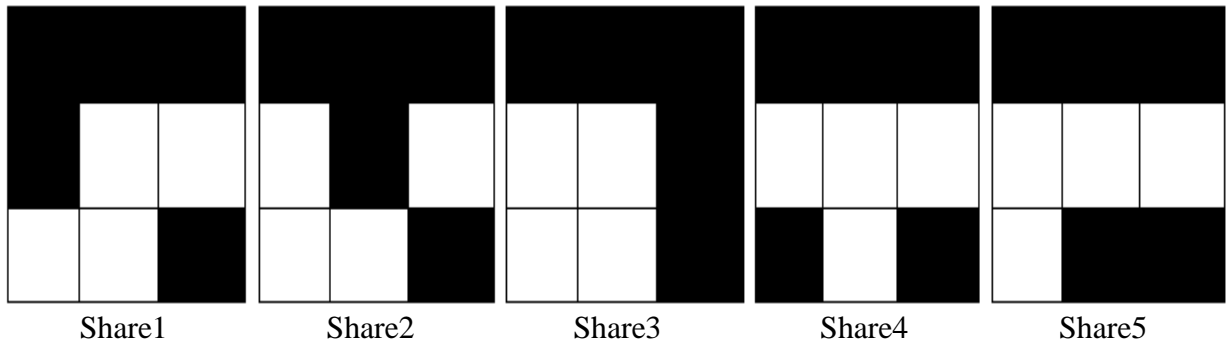

Figure 1. Partitions for black and white pixels

Superimposition of white and black pixels is shown below.

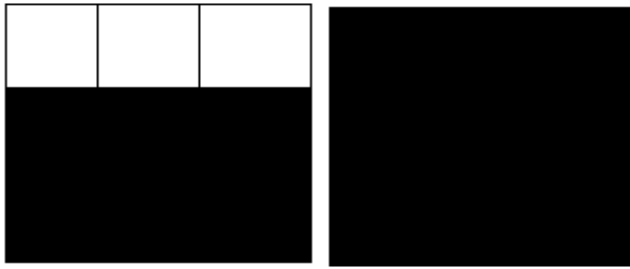

Any single share contains 4 black and 4 white pixels, so to make it a complete square array without distorting their aspect ratio, we need to add one more pixel. It should be either black or white.

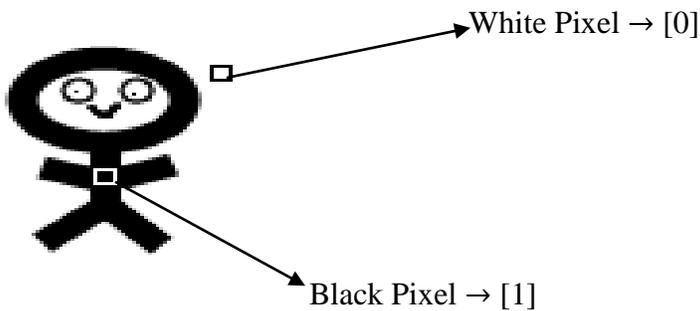

Figure 2. First Secret Image

Figure 2 is an example that shows how to transform the original pixels according to the shares we have designed; by this way we can maintain the aspect ratio of the image. These are the matrices for white and black pixels while we go for simulation.

$$\text{White Pixel} \rightarrow [0] \rightarrow \begin{bmatrix} 000 \\ 000 \\ 000 \end{bmatrix}_{3 \times 3} \qquad \text{Black Pixel} \rightarrow [1] \rightarrow \begin{bmatrix} 111 \\ 111 \\ 111 \end{bmatrix}_{3 \times 3}$$

## 3. RECURSIVE INFORMATION HIDING IN *3* OUT OF *5* SCHEME

Recursive information hiding is a technique where certain additional secret information can be hidden in one of the shares of the original secret image [4]. Here the secret information which we are going to hide is taken according to their sizes i.e. small images to larger. The first small secret image is divided into five different shares using visual cryptography. These shares are placed in the next level to create the shares of larger secret information. We are distributing the shares at each consecutive level so that no one has access to all the shares of the smaller images, unless until at least three participants come together to reveal the secret information, resulting in *3* out of *5* scheme.

Figure 3, is an example to help readers to understand the concept visually. The original secret image under consideration is of size 5 × 5 and the first secret image is of size 1 × 1. There are totally five shares obtained for the first secret image based on the basic idea of visual cryptography. The second secret image is of size 5 × 1. To obtain the second secret image we are going to use the shares of first secret image and they are placed in different levels (level 1-5) one after the other in five different shares. Now the shares of second secret image are designed by seeing the shares of first secret image and the original second secret image.

From Figure 3, we can see share 2 of the first secret image is placed on share 2 of the second secret image (level 2), and this share is a part of complete black pixel. The original second secret image has a complete white pixel under level 2, so we need to design all the shares of level 2 of second secret image by comparing both the shares of first secret image and the original second secret image to get a partial white pixel when we combine all the 5 shares or at least 3 shares in our case. This process is recursively repeated for all the shares. By doing this we can obtain the information of original second secret image by combining any 3 shares. This is the process for recursive information hiding of images. From Figure 1 we can see that the partitions of white pixel are stacked upon each other three fifth of the pixel is white and hence appears light gray to human eye. However, the subpixels of the black pixel are not complete black when 3 shares are stacked together but it is completely black when all the 5 shares are stacked.

Figure 4 shows an illustration of recursive information hiding. The original secret image considered is the Lena image of size 380 × 390. The first secret image is stick figure of size 78 × 78 and the second secret image is a text of size 380 × 78 and these both are hidden recursively.

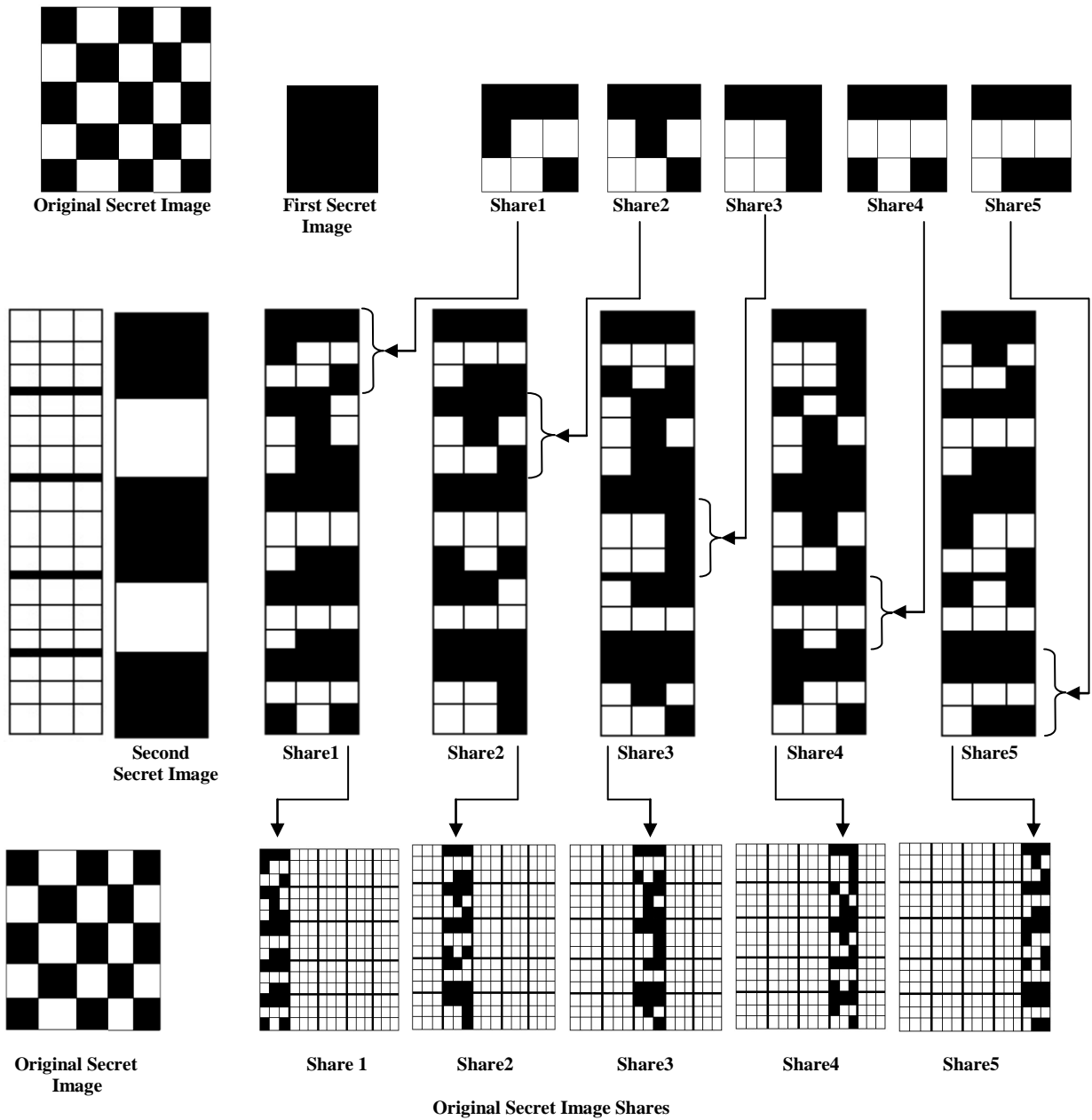

**Figure 3. Representation of Recursive Information Hiding of secret images in the shares of larger original image using a *3* out of *5* threshold scheme.**

Shares of original secret image are constructed by using the shares of second secret image and placed in different levels and the process gets repeated and finally when we combine any 3 shares the original secret information is revealed.

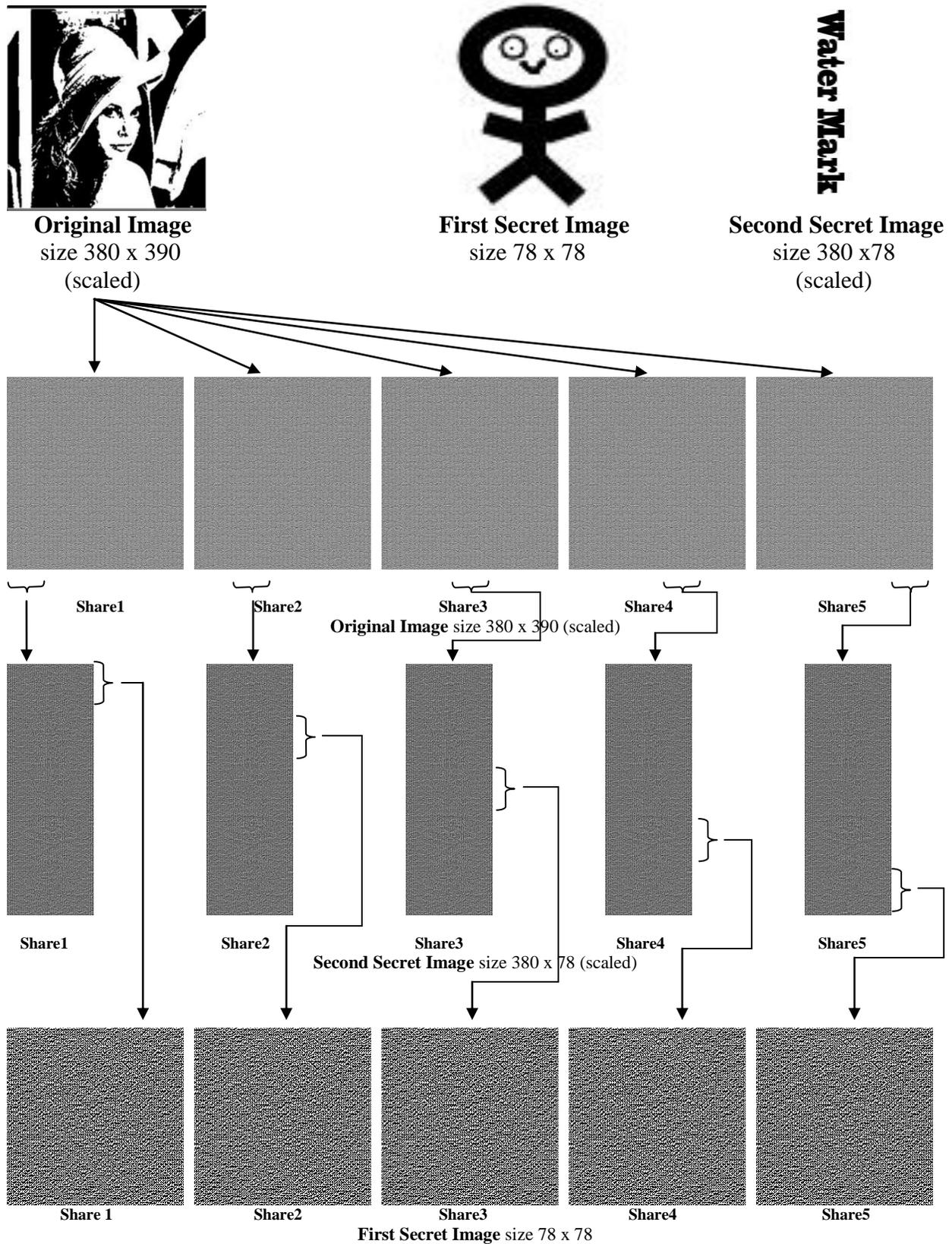

**Figure 4. Interpretation of the process of recursive information hiding of secrets in shares of larger original image**

After simulation these are few images than can reveal the secret information by combining any *3* shares out of *5* shares.

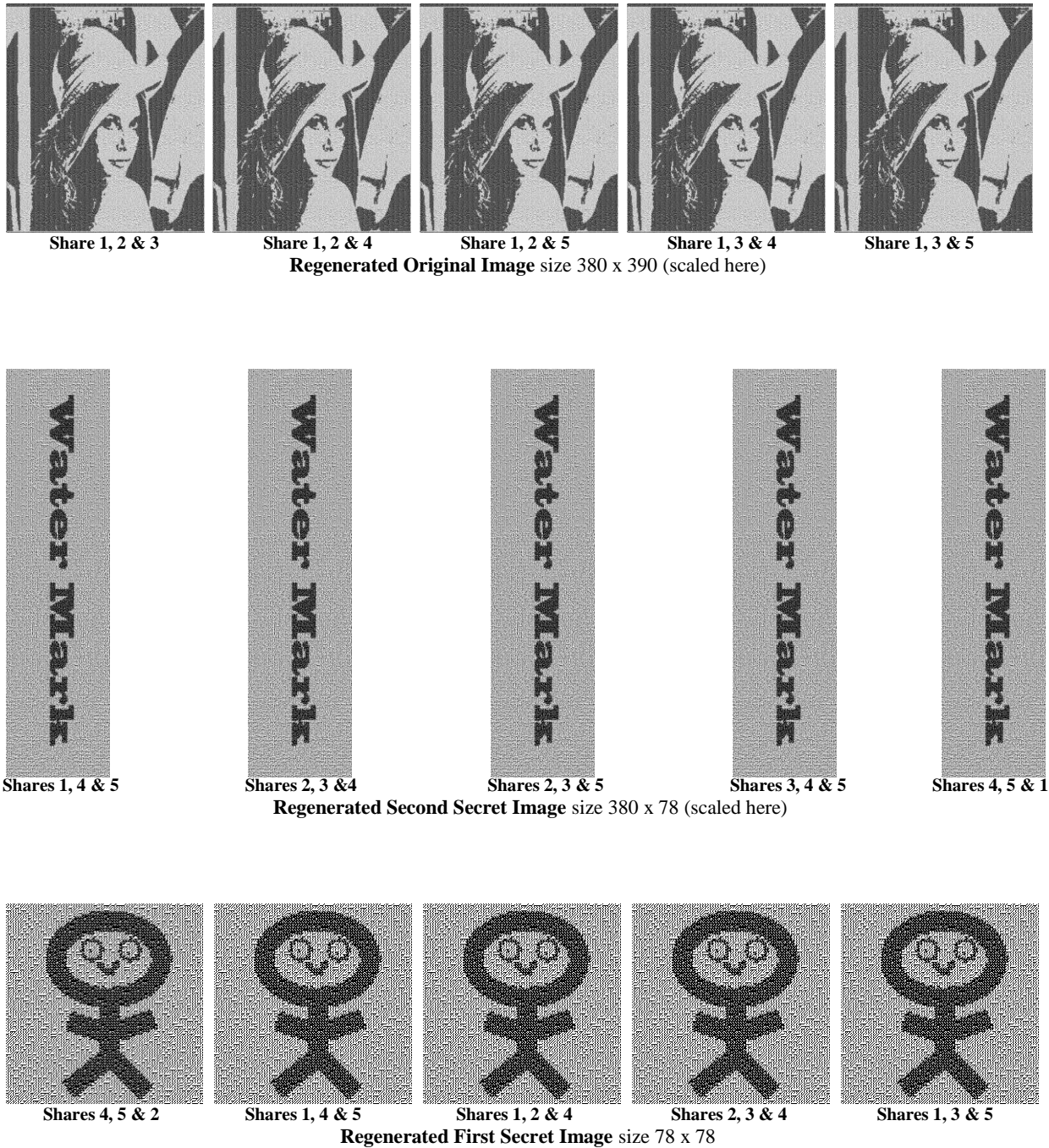

Figure 5. Regenerated smaller images from the shares hidden inside the shares of the original larger image.

## 4. A GENERAL "*k*" out of "*k*" SCHEME

For all $k$ there exists a general construction of $k$ out of $k$ visual secret sharing scheme, the pixel expansion must use at least $2^{k-1}$ pixels, and the relative contrast should be $\frac{1}{2^{k-1}}$. There is a need to construct two collections of $k \times 2^{k-1}$ Boolean matrices i.e. $S^0$ and $S^1$.

- $S^0$ Handles the white pixels.
- $S^1$ Handles the black pixels.

All $2^{k-1}$ columns have an even number of 1's in $S^0$ and odd number of 1's in $S^1$ and no two $k$ rows are same in both $S^0$ & $S^1$. $C_0$ and $C_1$ contains all permutations of columns in $S^0$ & $S^1$.

**Properties of "*k*" out of "*k*" scheme**

- Pixel Expansion, $m = 2^{k-1}$ ( $m$ should be as small as possible)
- Relative Contrast, $\alpha = \frac{1}{2^{k-1}}$ ( $\alpha$ should be as large as possible)
- $r$, the size of the collections. $C_0$ and $C_1$ (they need not be the same size, but in all of our constructions they are). Here $r = 2^{k-1}!$
- log$r$ is number of random bits needed to generate share.

By Naor and Shamir, any $k$ out of $k$ scheme as $\alpha \leq \frac{1}{2^{k-1}}$ and $m \geq 2^{k-1}$

Based upon the following properties we can design the matrix for *k = 3* and *k =4*

$k = 3$, therefore $m = 4$, α = ¼ and $r = 24$

$$S^0 = \begin{bmatrix} 0 & 0 & 1 & 1 \\ 0 & 1 & 0 & 1 \\ 0 & 1 & 1 & 0 \end{bmatrix}_{3 \times 4} \qquad S^1 = \begin{bmatrix} 0 & 0 & 1 & 1 \\ 0 & 1 & 0 & 1 \\ 1 & 0 & 0 & 1 \end{bmatrix}_{3 \times 4}$$

$k = 4$, therefore $m = 8$, α = $\frac{1}{8}$ and $r = 40320$

W = {1, 2, 3, 4}

Even cardinality subsets {{ }, {3, 4}, {2, 4}, {2, 3}, {1, 4}, {1, 3}, {1, 2}, {1, 2, 3, 4}}

Odd cardinality subsets {{4}, {3}, {2}, {2, 3, 4}, {1}, {1, 3, 4}, 1, 2, 4}, {1, 2, 3}}

$$S^0 = \begin{bmatrix} 0 & 0 & 0 & 0 & 1 & 1 & 1 & 1 \\ 0 & 0 & 1 & 1 & 0 & 0 & 1 & 1 \\ 0 & 1 & 0 & 1 & 0 & 1 & 0 & 1 \\ 0 & 1 & 1 & 0 & 1 & 0 & 0 & 1 \end{bmatrix} \begin{matrix} \rightarrow \\ \rightarrow \\ \rightarrow \\ \rightarrow \end{matrix} \begin{matrix} \text{Share1} \\ \text{Share2} \\ \text{Share3} \\ \text{Share4} \end{matrix}$$

<div align="center"><em>4 × 8</em></div>

Hamming weight of $S^0$ i.e. is White $H(V) = 7$

$$S^1 = \begin{bmatrix} 0 & 0 & 0 & 0 & 1 & 1 & 1 & 1 \\ 0 & 0 & 1 & 1 & 0 & 0 & 1 & 1 \\ 0 & 1 & 0 & 1 & 0 & 1 & 0 & 1 \\ 1 & 0 & 0 & 1 & 0 & 1 & 1 & 0 \end{bmatrix} \begin{matrix} \rightarrow \\ \rightarrow \\ \rightarrow \\ \rightarrow \end{matrix} \begin{matrix} \text{Share1} \\ \text{Share2} \\ \text{Share3} \\ \text{Share4} \end{matrix}$$

<div align="center"><em>4 × 8</em></div>

Hamming weight of $S^1$ i.e. is Black $H(V) = 8$

We need two collections of $4 \times 8$ Boolean matrices $C_0$ and $C_1$, contains all permutations of columns in $S^0$ and $S^1$, by these two matrices we can design the shares of black and white pixels. Similarly as above, we can apply recursive scheme for any $k$ out of $k$ scheme.

## 5. A GENERAL "$k$" out of "$n$" SCHEME

A general $k$ out of $n$ scheme is designed from $k$ out of $k$ scheme. Let $C$ be $k$ out of $k$ visual secret sharing scheme with parameters $m$, $r$, $\alpha$. The scheme $C$ consists of two collections of $k \times m$ Boolean matrices and $C_0 = T_1^0, T_2^0, \ldots . T_r^0$ and $C_1 = T_1^1, T_2^1, \ldots . T_r^1$. $H$ is a collection of $l$ functions $\forall h \in H, h : \{1 \ldots n\} \rightarrow \{1 \ldots k\}$. Let $B$ be the subset of $\{1 \ldots n\}$ of size $k$ and $\beta_q$ is probability that randomly chosen function $h \in H$ yields $q$ different values on $B$, $1 \leq q \leq k$.

We construct from $C$ and $H$ a $k$ out of $n$ scheme $C^|$ with parameters $m^| = m \cdot l$, $\alpha^| \geq \beta_k \alpha$, $r^| = r^l$

- The ground set is $V = U \times H$
- Each $1 \leq t \leq r^l$ is indexed by a vector $(t_1, t_2, \ldots t_l)$ where each $1 \leq t \leq r$.
- The matrix $S_t^b$ for $t = (t_1, t_2, \ldots t_l)$ where $b \in \{0, 1\}$ is defined as

$$S_t^b[i, (j, h)] = T_{tj}^b[h(i), j]$$

**Contrast**

Contrast should be $\geq \beta_k \alpha$

- $k$ rows is $S_t^b$, mapped to $q < k$ different values by $h$
- Hamming weight of OR of $q$ rows is $f(q)$
- Difference is $\alpha m$ white and black pixels occurs when $h$ is one and happens at $\beta_k$
- WHITE: $H(v) \leq l(\beta_k(d - \alpha m) + \sum_{q=1}^{k-1} \beta_q \cdot f(q))$
- BLACK: $H(v) \geq l(\beta_k + \sum_{q=1}^{k-1} \beta_q \cdot f(q))$

**Security**

Security properties of the "*k* out of *k*" scheme imply the security of "*k* out of *n*" scheme because we are using (*k,k*) scheme to create (*k,n*) scheme. The expected hamming weight of OR of *q* rows, $q < k$ is $l \sum_{q=1}^{k-1} \beta_q . f(q)$ irrespective of WHITE or BLACK pixel.

## 6. CONCLUSION

Recursive information hiding in visual cryptography can be applied to many applications in real and cyber world. The advantage is that the final decryption process is done by human visual system instead of complex computations. In this article we have presented a *3* out of *5* recursive hiding scheme that can be extended to a *k* out of *n* scheme.